\documentclass[epj-spec]{svjour}
\usepackage{graphics}

\usepackage{amsmath,amssymb}

\usepackage[cmtip,arrow]{xy}
\usepackage{pb-diagram,pb-xy}
\usepackage{pst-plot, pst-coil, pstricks}

\begin{document}
\title{Implementation of a medium-modified parton shower algorithm}

\author{N\'estor Armesto\inst{1} \and Leticia Cunqueiro\inst{1}$^,$\inst{2} \and Carlos A. Salgado\inst{1}
}                     
%
%
\institute{Departamento de F\'isica de Part\'iculas and IGFAE, Universidade de
  Santiago de Compostela,\\ 15706  Santiago de Compostela, Spain \and Istituto Nazionale di Fisica Nucleare, Laboratori Nazionali di Frascati, \\ I-00044 Frascati (Roma), Italy}
%
%
\abstract{We present a Monte Carlo implementation of medium-induced gluon radiation in the final-state branching process. Medium effects are introduced through an additive term in the splitting functions. We have implemented such modification within PYTHIA. 
 We show the medium effects on the hump-backed plateau, and the transverse momentum and angular distributions with respect to the parent parton. As expected, with increasing medium densities there is an increase (decrease) of partons with small (large)  momentum fraction, and angular broadening is observed. The effects on the transverse-momentum distributions are more involved, with an enhancement of low- and intermediate-$p_T$ partons and a decrease at large $p_T$, which is related to energy conservation, and to the lack of momentum exchange with the medium in our approach.}
%
%
\maketitle
\section{Introduction}
\label{intro}

Jet quenching - the suppression of particles with large transverse momentum produced in nucleus-nucleus collisions compared with the expectations from an incoherent superposition of nucleon-nucleon ones - is one of the most striking experimental observations at the Relativistic Heavy Ion Collider (RHIC) at Brookhaven National Laboratory \cite{rhic}. Its usual explanation is radiative energy loss (see the reviews \cite{rel}). Nevertheless, these measurements at RHIC suffer from a trigger bias - the requirement of a high transverse momentum particle in the event - which affects the production mechanism. This bias is very difficult to consider in analytical models as it comes strongly related to energy-momentum constraints, while most models have been developed within high-energy approximations. On the other hand, radiative energy loss is not the unique explanation for jet quenching, and several implementations of radiative energy loss exist \cite{Majumder:2007iu}.

To better characterize the produced medium and to distinguish among these different possibilities, both more differential observables at large transverse momentum and unbiased measurements like jets, are required. While RHIC is starting to look at some of these new opportunities \cite{jets}, the Large Hadron Collider \cite{lhc} will be the ideal place due both to the higher collision energy and to the characteristics of the detectors.
 
 Radiative energy loss implies a modification of the standard QCD radiation pattern. The proper tool for considering the QCD branching process in the final state with full energy-momentum conservation, is a Monte Carlo simulator.
 In spite of the fact that a
 probabilistic interpretation of radiation in a medium requires phenomenological
 assumptions, the practical advantages of a Monte Carlo are numerous. First,  
 it allows the access to other observables different from the limited single
 inclusive measurements, such as different jet
 shapes, jet multiplicities, multiparticle intrajet correlations,$\dots$ Moreover, such an implementation makes it possible to explore new physical
 mechanisms in jet development, such as the interplay of the multigluon
 radiation with the medium length, effects of the color flow and
 reconnections, effects of recoil with the medium, etc. Several implementations of radiative energy loss in Monte Carlo codes exist \cite{mc}, but in them either the radiation process is superimposed on other effects like collisional energy loss, or it is treated in a simplified way as an enlargement of the QCD evolution or through a multiplicative increase of the collinear parts of the splitting functions \cite{BorghiniWiedemann}.

 In this paper we present a Monte Carlo with medium-modified final-state radiation, based on the
 ideas described in \cite{msf}. There, medium effects enter as an
 additive correction to the standard, vacuum splitting functions:
 \begin{equation}
P_{\rm tot} (z)= P_{\rm vac} (z)\to
 P_{\rm tot} (z)=P_{\rm vac}(z)+\Delta P(z,t,\hat{q},L,E).
 \label{eq1}
 \end{equation}
 This correction, $\Delta P(z,t,\hat{q},L,E)$ which we write as $\Delta P(z,t)$ in the following, depends not only on the energy fraction
 $z$ but also on the virtuality $t$ of the radiating parton and its energy $E$, and on the medium characteristics relevant for radiative energy loss: transport coefficient $\hat{q}$ and  medium length $L$.
 This simple medium modification is 
 implemented in the standard final-state showering routine PYSHOW in PYTHIA \cite{Sjostrand:2006za}.
 In our approach, the inelastic energy loss and the angular broadening of the
 shower are dynamically related through a single parameter, the transport
 coefficient $\hat{q}$. The longitudinal evolution of the shower is implemented  by considering the formation time of the radiated gluons.
 However, in its present form our  implementation does not consider either the recoil of the
 scattering centers - and consequently elastic energy loss is not taken
 into  account -, or the exchange of color with the scattering centers in the medium.

In this note we focus on the description of the implementation of medium-induced gluon radiation in the $t$-ordered final-state-radiation routine in PYTHIA, and present some results. For a discussion of the basis of the modification of the splitting functions \eqref{eq1}, we refer the reader to \cite{msf}.
Extensive discussions together with a comparison between medium effects on branching processes for different ordering variables, will be published elsewhere \cite{ourpaper}. A final version of our routine, called Q-PYTHIA, will be soon made publicly available \cite{ourpaper}.
 

\section{The Monte Carlo}
\label{montecarlo}

\subsection{Basics steps}

Basically, a branching algorithm must solve the following problem: given a
parton coming from a branching (or production) point with coordinates ($t_{1}$,$x_{1}$), with  $t_{1}$ the virtuality and
$x_{1}$ its energy fraction, which are the coordinates ($t_{2}$,$x_{2}$) for the next branching? 

Ignoring, for simplicity, parton labels, the Sudakov form factor 
\begin{equation}
\Delta(t_{1})=\exp{\left[-\int_{t_0}^{t_{1}} {dt^\prime \over t^\prime}
\int_{z^{-}}^{z^{+}} dz {\alpha_s(t_{1})
\over 2 \pi} P(z)\right]}
\label{eq2}
\end{equation}
 gives the probability for a parton not
to branch while evolving from an initial scale $t_{0}$ to another scale
$t_{1}$. Consequently, $\Delta(t_{2})/\Delta(t_{1})$ stands for the
probability of evolving from $t_{1}$ to $t_{2}$ without branching. Thus
$t_{2}$ can be generated by solving the equation
\begin{equation}
 \frac{\Delta(t_{2})}{\Delta(t_{1})}=R ,
 \label{eq3}
 \end{equation}  
$R$ being a random number.

The energy fraction kept by the parton in the next branching $z_{2}$ can be
diced down by solving the equation:
\begin{equation}
\int_{z_{-}}^{z_{2}} dz \frac{\alpha_{S}}{2\pi} P(z)=R^\prime
\int_{z_{-}}^{z_{+}} dz \frac{\alpha_{S}}{2\pi} P(z), \label{eq4}
\end{equation}
with $R^\prime$ another random number.
\eqref{eq3} and \eqref{eq4}  are the two basic steps of a branching Monte Carlo algorithm. 
 The only difference between the standard case and the one considering medium effects comes through the substitution \eqref{eq1}.
 
In all equations above, the value of the maximum initial virtuality, $t_{max}=4E_{jet}^2$, of the lower virtuality limit, $t_0=1$ GeV$^2$, the lower and upper limits of the $z$-integrals, $z_\pm\equiv z_\pm(t)$, and the scale at which $\alpha_s$ runs, as well as all other aspects of the $t$-ordered evolution, are the PYTHIA defaults, see \cite{Sjostrand:2006za}. These aspects will be described and discussed more extensively elsewhere \cite{ourpaper}.

\begin{figure}[h]
\resizebox{1.\textwidth}{0.4\textheight}{%
  \includegraphics{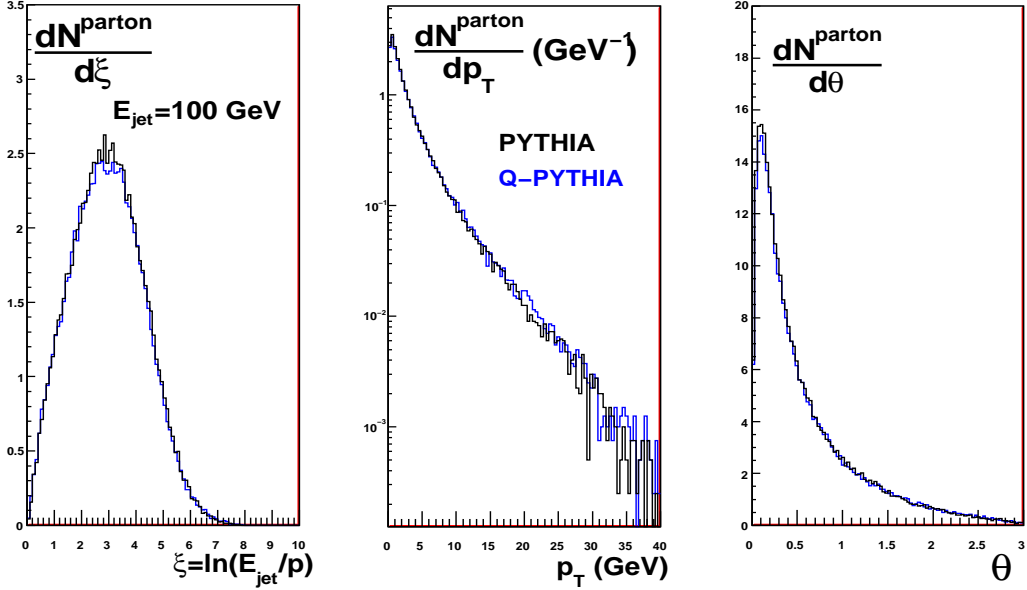}
}
\caption{Intrajet parton distributions in $\xi=\ln{(E_{jet}/p)}$ (left), $p_{T}$ (middle) and $\theta={\rm acos}(p_z/p)$ (right)
   for a gluon of initial energy $E_{jet}=100$ GeV. PYTHIA default (black lines)  and
   our results with $\hat{q}=0$ (blue lines) are compared.}
\label{fig:1}       
\end{figure}

PYTHIA approximates the splitting functions by their $z \to 1$ forms, and later this approximation is corrected via a rejection method. On the other hand, our splitting functions also include small $z$ corrections - we use for the vacuum the exact splitting functions.  This is the main source of the very small
differences between  our vacuum implementation ($\hat{q}=0$) and PYTHIA
default shown in Fig. \ref{fig:1}. We take this comparison as a check of the quality of our procedure.
Note also that our medium modifications are for $g\to gg$ and $q(\bar q)\to q(\bar q)g$
splittings \cite{msf}. The $g\to q\bar{q}$ branching is the same as in vacuum because
its splitting probability is not singular at $z\to 1$. 

\subsection{Length and energy evolution}
\label{seclen}

The length traveled by a parton before a gluon decoheres from its wave
function and is radiated, can be estimated \cite{rel} by the
gluon formation length $l_{coh}={2 \omega/k_{T}^2}$, where $\omega$ and
$k_{T}$ are the energy and transverse momentum (with respect to the parent parton)
of the emitted gluon,
respectively.   

The shower begins with a parton that faces the full length of the medium $L$, so
the medium effects on the probability of the first branching are evaluated at $L$. The coherence
length of the emitted gluon is then computed being its next branching
evaluated at $L-l_{coh}$. The process is iterated.

Also the energy degradation is considered. For a process $a(E_b+E_c)\to b(E_b)+c(E_c)$, the medium effects in the branching process of $a$ is considered at energy $E_b+E_c$, while the subsequent branchings of $b$ and $c$, if any, are considered at $E_b$ and $E_c$ respectively. In our default results for the medium, both the evolution in length and the energy degradation are considered. The separate effect of these aspects of evolution will be discussed elsewhere \cite{ourpaper}.

\section{Results}

At the parton level , the general expectations of medium-induced gluon radiation formalisms are:
\begin{itemize}
  \item A softening of the spectra.
  \item An increase of the parton multiplicity. 
  \item An angular broadening of the jet.
\end{itemize}

\begin{figure}[h]
\resizebox{1.\textwidth}{0.4\textheight}{%
  \includegraphics{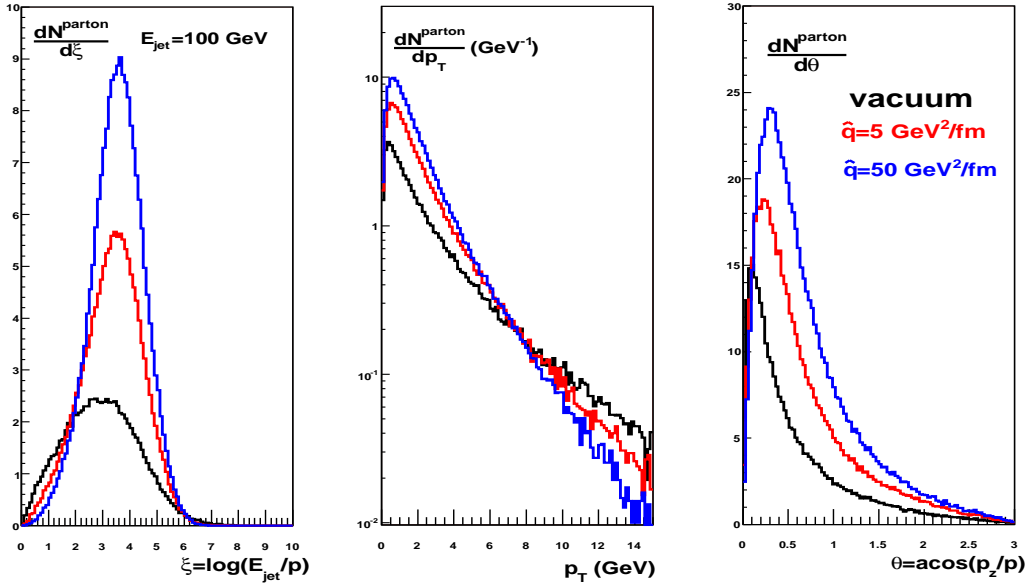}
}
\caption{Intrajet parton distributions in $\xi$ (left), $p_{T}$ (middle)
 and $\theta$ (right) for a gluon of initial energy $E_{jet}=100$ GeV in a medium of
 length  $L=2$ fm and for different transport coefficients $\hat{q}=0$ (black), 5 (red) and 50 (blue lines) GeV$^2$/fm.}
\label{fig:2}       
\end{figure}

To see the results, shown in Figs. \ref{fig:1}-\ref{fig:2}, we run PYTHIA final-state-radiation routine PY\-SHOW with our medium modifications, on a gluon of
energy $E_{jet}=100$ GeV moving along the positive $z$-axis, and study the intrajet distribution of
final partons in energy fraction (actually in $\xi=\ln{(E_{jet}/p)}$ with $p=|\vec{p}|$ the modulus of the momentum of the final particle - the hump-backed plateau plot), transverse momentum $p_T=\sqrt{p_x^2+p_y^2}$ of the final particle, and polar angle $\theta={\rm acos}(p_z/p)$ of the final particle. The medium length is fixed to $L=2$ fm.
For this note, we focus on representative results which illustrate the effects of medium-induced gluon radiation in the branching process. The statistics we have used is $10^5$ generated events. Further results will be published elsewhere \cite{ourpaper}.

In Fig. \ref{fig:2} we show the results at parton level, for two values of the transport coefficient $\hat{q}=5$ and 50 GeV$^2$/fm. We observe a suppression of high-$z$ particles and a large
enhancement of particles with low-intermediate $z$-values, as expected.
We also observe a suppression of high-$p_{T}$
particles and the corresponding enhancement of intermediate-$p_{T}$ particles.
The $p_T$-spectrum should be softer than vacuum at low transverse momentum since low-$p_{T}$ particles should be  kicked towards
higher values of the transverse momentum. However we find a clear enhancement
of low-$p_{T}$ particles. Here, the lack of exchange of energy and momentum
with the medium plus energy conservation in PYTHIA, is making a large effect. 
Finally, we see that the angular distribution broadens with increasing transport
coefficient, as  expected.

\section{Conclusions and outlook}

In this note we have presented an implementation of medium-induced gluon radiation in the final-state branching process. Medium effects are introduced through an additive term in the splitting functions. We have implemented such modification within PYTHIA \cite{Sjostrand:2006za}. In this note, we focus on a case study. Extensive discussions of the proposed formalism, together with a comparison between medium effects on branching processes for different ordering variables, will be published elsewhere \cite{ourpaper}. The corresponding implementation in HERWIG will be also presented there and computer codes will be released for public use.
 
In this brief discussion, we have observed the appearance of the different medium effects on the hump-backed plateau, and the transverse momentum and angular distributions with respect to the parent parton. As expected, with increasing medium densities there is an increase (decrease) of partons with small (large)  momentum fraction, and angular broadening  is observed. The effects on the transverse-momentum distributions are more involved, with an enhancement of low- and intermediate-$p_T$ partons and a decrease at large $p_T$, which is related to energy conservation and to the lack of momentum exchange with the medium in our approach.

Let us mention several caveats of our approach:
\begin{itemize}
\item A Monte Carlo implementation of medium-induced gluon radiation assumes that
there is an ordering variable in medium, this not having been theoretically proved yet.
\item As mentioned, there is no energy and momentum exchange with the medium in our
present implementation. This could be of importance in an energy
conserving Monte Carlo such as PYTHIA. 
\item Elastic energy loss is not considered. 
\item The effects on the color flow between the jet and the medium have to be considered.
\end{itemize}
Together with the possibility of studying the medium modifications in a realistic
environment using state-of-the-art reconstruction and pileup subtraction techniques,  these caveats constitute our to-do list.

\section*{Acknowledgments}

We thank N. Borghini, D. d'Enterr\'{\i}a, F. Krauss, T. Sj\"ostrand and U. A. Wiedemann for useful discussions. Special thanks are due to G. Corcella for an ongoing collaboration.
This work has been supported by Ministerio de Educaci\'on
y Ciencia of Spain under project FPA2005-01963, by Xunta de Galicia (Conseller\'{\i}a de Educaci\'on), and by the Spanish Consolider-Ingenio 2010 Programme CPAN (CSD2007-00042). 
NA has been supported by  MEC of Spain under a contract Ram\'on y Cajal, and Xunta de Galicia through grant PGIDIT07PXIB206126PR. CAS has been supported by  MEC of Spain under a contract Ram\'on y Cajal and by the European Commission grant PERG02-GA-2007-224770.

%

%
%

\end{document}